          [2001/04/25 1.1 (PWD)]
\documentclass[a4paper,twocolumn]{esapub} % European paper
\usepackage{natbib}

\title{Primordial Resonant Lines and Observational Strategy for ODIN and HERSCHEL}
\author[1]{R. Maoli}
\affil[1]{Dipartimento di Fisica, Universit\`a di Roma, ``La Sapienza'', 
Piazzale Aldo Moro, 2, I-00185 Roma, Italy}
\author[2]{P. Chambaud}
\affil[2]{Laboratoire de Chimie Theorique, Universit\'e de Marne la Vall\'ee, 
Champs sur Marne, France}
\author[3]{J.Y. Daniel}
\affil[3]{Institut d'Astrophysique de Paris, CNRS, 98 bis Boulevard Arago, 
75014 Paris, France}
\author[1]{P. de Bernardis}
\author[4]{P. Encrenaz}
\affil{LERMA, Observatoire de Paris, 61, Av. de l'Observatoire, 75014 Paris, 
France}
\author[1]{S. Masi}
\author[1]{B. Melchiorri}
\author[1]{F. Melchiorri}
\author[4]{L. Pagani}
\author[2]{P. Rosmus}
\author[4]{M. Signore}

\begin{document}


\maketitle

\begin{abstract}
The presence of a primordial molecular or atomic medium can give us the possibility to investigate 
the first phases of structure formation. The lines produced by resonant scattering of the CMB 
photons are the most important signals coming from the {\it dark age} of the post--recombination 
universe. The possibility to detect these lines with ODIN and Herschel satellites is investigated. 
\end{abstract}

\section{Introduction}
When the temperature of the universe drops below 3000 K, the atoms of 
hydrogen recombine and the universe becomes transparent to the cosmic 
microwave background (CMB) photons.
A {\it dark age} from the observational point of view starts. This epoch lasts 
till the universe is reionized and photons--electrons interaction becomes 
effective again.

Following this scenario, the observed signal coming from primordial 
structures can be divided in three classes: 
\\
- {\bf CMB primary anisotropies}: directly related to the matter perturbations 
at $z\sim 1100$;
\\
- {\bf CMB gravitational secondary anisotropies}: due to the interaction 
between the CMB photons and the gravitational field at $z < 1100$ 
(Rees--Sciama effect, integrated Sachs--Wolfe and gravitational lensing);
\\
- {\bf CMB post-reionization secondary anisotropies}: due to the interaction 
between the CMB photons and the newly available electrons (Vishniac and 
Sunyaev--Zeldovich effects).

In order to complete this list, we have to add interactions between the CMB 
photons and different chemical species (atoms, ions and molecules) that 
populate the cosmological medium. 
The abundance of these elements is not comparable to that one of free electrons
in the reionized universe, nevertheless the interaction through resonant 
scattering is more efficient than Thomson scattering resulting in the 
production of a different kind of secondary anisotropies: {\bf primordial 
resonant lines}.

The role of molecules in the early universe was first suggested by \citet{Dubrovich77}. The 
consequences on the CMB anisotropies were described by Maoli and collaborators 
(\citeauthor{Maoli94} 1994, \citeauthor{Maoli96}, 1996) with a first observational campaign at the 
30 meter IRAM radiotelescope \citep{debe93}. Recently the same mechanism has been applied to atoms 
and ions \citep{Basu04}.

The primordial cosmological medium can dump CMB primary anisotropies and produce lines through moving
scattering sources in the pre--reionization universe as well as in the post-reionization one.

\section{Chemistry of the primordial universe}

In order to quantify the effect of primordial resonant lines we have to investigate which kind of 
chemical species presumably populated the cosmological medium at high redshifts.
Assuming a homogeneous Standard Big Bang Nucleosynthesis model, we have a different composition for 
the pre--reionization cosmological medium and the post--reionization one: the former has atoms, 
ions and molecules issued from the light elements formed during the primordial nucleosynthesis, 
while the latter can have also chemical species from heavier elements.

\subsection{Primordial chemistry}

The chemistry of the pre--reionization universe follows the primordial nucleosynthesis when the 
temperature of the universe decreased enough to allow chemical reactions.
In the standard Big Bang model the light elements (hydrogen, deuterium, helium and lithium) are 
synthesized in the very early universe when $T\sim 0.1$ MeV and $t\sim 100$ s.

The chemistry of these light elements begins around the recombination epoch. The final results are 
very uncertain due to the high number of the involved chemical reactions and the difficulty to 
compute or to measure all the reaction rates in the typical conditions of the early universe
\citep{Puy02}.
For example the final abundances for LiH molecules can be different by six orders of magnitude 
following different papers (\citeauthor{Lepp84}, 1984, \citeauthor{Stancil96}, 1996).

\subsection{Chemistry of heavy elements}

Heavy elements are synthesized in possible population III stars and later in population II stars 
(mainly during the later stages of their evolution i.e. when they become supernovae).

When quasars appeared in the universe they are believed to have reionized the medium. For long time 
quasars with a break at the hydrogen reionization energy has been searched in order to prove that 
there is some intervening neutral material. Is is the well known Gunn Peterson (GP) test.

While it is clear that quasars provide a significant contribution to the ionising background at low 
redshift ($z\sim$ 3), it is difficult to think that they are capable to maintain the ionized state 
of the cosmological medium at $z=6-8$.
Moreover, a possible explanation for the wide spread abundances of heavy elements in the 
intergalactic medium today can be provided by an early star formation dominated by very massive 
stars (the so called population III stars). This population adds its contribution to that one of later 
population II stars.
Until recently there was a consensus about a relatively late ionization epoch which was suggested 
in particular by the observation of a full GP trough in a quasar at $z$=6.28 in the SDSS 
\citep{Becker01}.

This was the situation until the release of the first year of WMAP data \citep{Kogut03}. In these 
data the high optical depth to Thomson scattering ($\tau \sim 0.17 \pm 0.04$) implies an ionization 
redshift of $17\pm 3$ under the assuption that reionization occurs abruptly.

A priori there are no simple reionization models which are consistent with the combination of the 
WMAP results and data from the low redshift universe ($z \leq 6$). Reionization clearly does not 
occur through a single rapid phase transition. At least one must modify the simplest model by 
adding a second transition.

A lot of scenarios appear in the last year. Anyway, after WMAP, there is evidence for an early 
generation of stars existing 4 to 5 times earlier than any object yet observed.

Heavy elements can play a major role in the medium composition starting from $z=20-25$.

\section{Interaction between the CMB photons and the primordial cosmological 
medium}

Being the primordial medium non collisional and in equilibrium with radiation, 
absorption and emission processes are ineffective and resonant elastic 
scattering must be considered as the most efficient in coupling matter and 
radiation.
In this process, a photon is absorbed by an atom, an ion or a molecule at a 
frequency $\nu_{ij}$ corresponding to a line transition of the considered 
chemical species and reemitted at the same frequency.

The cross section for this process is given by the formula:
\begin{equation}
\sigma_{res}=\frac{\lambda_{ij}^3 A_{ij}}{4c}|\vec{d}|^2
\frac{\nu_{ij}}{\Delta\nu}=3.37\cdot 10^{-19}|\vec{d}|^2
\frac{\nu_{ij}}{\Delta\nu}
\end{equation}
where $\lambda_{ij}$ and $\nu_{ij}$ are the wavelenghts and the frequency of 
the transition, $\Delta\nu$ is the broadening of the line (Doppler broadening 
in the non collisional primordial medium), $\vec{d}$ is the dipole moment of 
the species in Debye. 

For a dipole moment of order of unity and typical line 
widths $\frac{\Delta\nu}{\nu}=10^{-4}- 10^{-6}$, the resonant scattering is
$10^{10}- 10^{12}$ times more efficient than Thomson scattering 
($\sigma_T=6.652\cdot 10^{-25}\, {\rm cm}^2$). 
This explains why the coupling with 
atomic or molecular medium can be stronger than the coupling with electrons 
in spite of the low abundance of the former.
With a dipole moment of 5.88 Debye, the LiH molecule is a very good candidate 
for an effective interaction with CMB photons. Unfortunately its abundance 
seems to be very low.

Being the resonant scattering an elastic process, it can result in a CMB primary anisotropy 
attenuation and secondary anisotropy production, where chemical species play exactly the same role 
as electrons in the reionized universe.
Smearing of the primary anisotropies is very small but could be marginally detected by future CMB 
experiments using its peculiar frequency dependence behaviour \citep{Maoli94}.
Secondary anisotropies raise if the scattering source has a non zero component of the peculiar 
velocity along the line of sight. In this case the elastic scattering is no more isotropic in the 
observer frame.

The strong frequency dependence of the cross section is responsible for the differences between 
molecular (or atomic) secondary anisotropies and those one produced by Thomson scattering in an 
early reionized universe: we have molecular (or atomic) primordial lines instead of fluctuations 
of the cosmic black body temperature. These lines can be emission--like or absorption--like 
depending on the sign of the radial peculiar velocity of the scattering source.

The signal is given by the formula:
\begin{equation}
\frac{\Delta I}{I_{CMB}}=\left(1-e^{-\tau_{ij}}\right) \beta_{p,r}\left(3-\alpha_\nu\right)
\end{equation}
where $\beta_{p,r}$ is the radial component of the peculiar velocity normalized to the speed of 
light, $\alpha_\nu$ is the spectral index of the photon distribution, and $\tau_{ij}$ is the 
optical depth of the scattering source computed at the observational frequency:
\begin{equation}
\tau_{ij}=\int_{source}{\sigma_{res,ij}n_ic\,dt}
\end{equation}
where $n_i$ is the numerical density of the species in the state $i$.

Detection of primordial resonant lines is comparable to Lyman--$\alpha$ cloud detection using 
radiation coming from high redshift quasars. In our case, Lyman--$\alpha$ clouds are substituted by 
primordial structures, while the role of quasar is played by the CMB. For primordial lines, 
absorption can't be efficient because of the temperature of the CMB, so only elastic processes can 
be considered. As a consequence, also a peculiar velocity is needed to produce the signal.

\section{Observational strategy for primordial resonant lines detection}

\subsection{Which kind of instrument?}

Due to the spectral behaviour of the resonant scattering cross section, what we are looking for are 
spectral lines. For a fixed observational frequency, $\nu_{obs}$, the source producing the line 
must be only at a fixed redshift given by the formula: 
\begin{equation}
z_s=\frac{\nu_{ij}}{\nu_{obs}}-1
\end{equation}
where $\nu_{ij}$ is the frequency of the line transition.
As an example, considering the LiH molecule, its first five rotational states, and 
$\nu_{obs}=100$ GHz, only sources at $z_s$=3.44, 7.87, 12.29, 16.70 and 21.08 can give a 
contribution to the signal.
Now, the probability to have, along the line of sight more than one source, at these given 
redshifts, with the dynamical conditions (peculiar velocity, molecular or atomic abundance) 
necessary to produce the signal, is very low.

As a consequence, we have not to consider signal superposition but we look for a single spectral 
line produced by a well identified primordial source. The detection of this line will give us the 
indication of the existence of a protostructure at a given redshift. The properties of the line 
will provide us indications on the mass, peculiar velocity and chemical composition of this source, 
allowing us to constrain structure formation models.

A high sensitivity spectrometer working in the millimeter and submillimeter range is needed.

\subsection{Band width}

Improving the band width will improve the explored redshift range 
$\left(\frac{\Delta z}{z}=\frac{\Delta\nu_{obs}}{\nu_{obs}}\right)$ and the probability to detect a 
primordial line will be higher. 
At a certain point the redshift range explored by a line will superpose the redshift range of the 
next line and we will be sensible to the signal produced by a source at any redshift. The band 
width necessary to fulfill this condition depends on the considered chemical species and the 
observational frequency.

For example, to explore a redshift range between 10 and 25, considering the LiH molecule, we need a 
band with of 40 GHz for $\nu_{obs}=100$ GHz as well as for $\nu_{obs}=500$ GHz. 
For the lowest frequency this is certainly a technical challenge but the expected signal is higher 
because low rotational transitions are concerned, while for the highest frequency this band is 
easier to be obtained but the signal will be lower being produced by high rotational lines.

Another advantage of a broad band width is the possibility to detect two different lines produced 
by the same molecule (or atom) and the same source. This is the best way to check the origin of the 
line.

\subsection{Angular and spectral resolution}

Computation has been done for the line amplitude, the line width and the angular size typical of a 
molecular signal produced during the linear evolution, the turn--around phase and the beginning of 
the non linear collapse of a proto--structure.

During the linear evolution of the primordial cloud, the perturbation follows the expansion of the 
universe; different parts of the cloud fulfill the resonant condition for different redshifts. The 
line width therefore depends on the extension in redshift of the perturbation, i.e. on its mass and 
redshift.
Primordial lines produced in this evolutionary phase will be very broad: $\frac{\Delta\nu}{\nu}$ 
can be as high as 0.01 for $10^{14}\, M_\odot$ perturbations at high redshifts.
Unfortunately the intensity is very low and the detection is penalized comparing with primordial 
lines produced in a more advanced evolutionary phase.

When the primordial cloud stops following the expansion of the universe (turn--around), all the 
scattered photons contribute to the signal at the same frequency. The line width is due to the 
thermal broadening and is very small (a few in $10^{-6}$ for $\frac{\Delta\nu}{\nu}$). Sources at 
turn--around produce the strongest and thinnest primordial lines.

During the non-linear collapse two different velocities have to be considered: the peculiar 
velocity $\beta_p$ of the cloud and its infall velocity $\beta_c$.
Using a simple model of spherical collapse, only a strip of the cloud having the same projection of 
the infall velocity along the line of sight will contribute at the same observational frequency. 
The line width is roughly twice the maximal infall velocity and the line shape depends on the ratio 
$\frac{\beta_p}{\beta_c}$; if $\beta_p$ is negligible, we have a double peak line.

This is the only signal emitted by a no moving primordial cloud and it is the consequence of the 
strong frequency dependence of the resonant scattering cross section; in the same situation the 
Thomson scattering will not produce any signal.

The double peak feature progressively disappears when the peculiar velocity becomes more important. 
Typical line width is $10^{-5}- 10^{-3}$ in $\frac{\Delta\nu}{\nu}$.

Angular sizes of a primordial cloud at turn--aroud or during the non--linear collapse can range 
from 10 arcseconds to a few arcminutes.

\subsection{Foreground contamination}

CMB measurements are limited by the presence of foregrounds. 
Galactic dust emission, bremstrahlung and synchrotron as well as thermal S--Z effect from clusters 
and high redshift thermal emission by dust are all possible sources of contamination.
Multi--frequency experiments are needed to compute the foreground contribution and to extract the 
CMB signal.

Primordial resonant line observations are completely free from this kind of problem: due to the 
resonant scattering spectral nature, all these foregrounds will represent a small, completely 
negligible, perturbation of the baseline.
Lines produced by the interstellar medium or extragalactic sources can be easily identified by 
looking for other lines of the same chemical species.

\subsection{Observational situation}

First searches for primordial resonant lines were performed few years ago using the 30 meter IRAM 
telescope at Pico Veleta (Spain). Observations were performed at three different frequencies in the 
1.3mm, 2mm and 3mm atmospheric windows. They were characterized by a small band width 
($2\times 500$ MHz) and concerned only five spots in the sky. In this case we obtained only upper 
limits for the abundance of the considered chemical species as a function of the redshift for small 
interval in $z$ and in a given direction.

New observations are planned using ODIN satellite. This is a Swedish satellite, developed in 
collaboration with France, Finland and Canada, launched in 2001 and equipped with a 1.1 meter 
telescope.
Two sky regions will be observed between 490 and 520 GHz with a resolution of 1 MHz and a 
sensitivity of 1 mK over 10 MHz.
The great improvement of these observations is the broadness of the band, available only for space 
experiments because of the atmospheric absorption.
In this way a wide range of redshifts will be explored.

The future for primordial resonant lines is represented by Herschel, the ESA satellite, that will 
allow to observe a statistical representative area in the sky with a large band width and high 
sensitivity.

\section{Conclusions}

Primordial resonant lines are probably the main tool to explore the pre--reionization and the 
post--reionization universe. The very rich information associated with spectral lines can help to 
understand the formation and evolution of structures as well as the history of the metal enrichment 
of the high redshift cosmological medium.

In this paper we concentrated to the observational strategy to detect these lines produced by 
nearly formed structures. In this case the spectral nature of the signal makes it almost free from 
galactic and extra--galactic contaminations.

Signals associated with a linear evolution phase of perturbations are weaker and broader and their 
detection would be possible, as well as the primary anisotropy smearing, with millimetric multifrequency 
photometers under the condition of a 
very high sensitivity and a very precise characterisation of foregrounds. 
This last requirement seems to be very challenging.


\begin{thebibliography}{}
\bibitem[Basu et~al.(2004)]{Basu04}
Basu K., Hern\'andez--Monteagudo C., Sunyaev R., 2004, A\&A 416, 447

\bibitem[Becker et~al.(2001)]{Becker01}
Becker R.H., et~al., 2001, Astron. J. 122, 2850

\bibitem[de Bernardis et~al.(1993)]{debe93}
de Bernardis P., et~al., 1993, A\&A 269, 1

\bibitem[Dubrovich(1977)]{Dubrovich77}
Dubrovich V.K., 1977, Soviet Astron. Lett. 3, 128

\bibitem[Kogut et~al.(2003)]{Kogut03}
Kogut A., et~al., 2003, ApJS 148, 161

\bibitem[Lepp and Shull(1984)]{Lepp84}
Lepp S., Shull M.J., 1984, ApJ 280, 465

\bibitem[Maoli et~al.(1994)]{Maoli94}
Maoli R., Melchiorri F., Tosti D., 1994, ApJ 425, 372

\bibitem[Maoli et~al.(1996)]{Maoli96}
Maoli R., et~al., 1996, ApJ 457, 1

\bibitem[Puy and Signore(2002)]{Puy02}
Puy D., Signore M., 2002, New Astron. Rev. 46, 709

\bibitem[Stancil, Lepp and Dalgarno(1996)]{Stancil96}
Stancil P.C., Lepp S., Dalgarno A., 1996, ApJ 458, 401

\end{thebibliography}
\end{document}